\documentclass[a4paper]{article}

\usepackage{amsmath}
\usepackage{amssymb}
\usepackage{latexsym} 
\usepackage{url}
\usepackage{bussproofs}
\usepackage{multicol}

\usepackage{tikz}
\usetikzlibrary{positioning}
\usetikzlibrary{plotmarks}

\newcommand{\squaredots}{::}
\newcommand{\Coloneqq}{::=}
\newcommand{\ldot}{\,.\,}

\newcommand{\lam}{\lambda\,}
\newcommand{\piTy}{\Pi\,}
\newcommand{\piAbs}[2]{\Pi\,#1\,.\,#2}
\newcommand{\infer}{\,: _{\uparrow}}
\newcommand{\tycheck}{\,: _{\downarrow}}
\newcommand{\bdot}{\,.\,}

\newcommand{\named}[1]{\mathtt{#1}}
\newcommand{\subst}[2]{[#1 \mapsto #2]}

\newcommand{\idTy}[3]{\mathrm{Id}_{#1}(#2, #3)}
\newcommand{\refl}{\mathrm{refl}}

\newcommand{\sysname}{\textsc{Mella}}

\newcommand{\elimJ}{\mathrm{elimJ}}

\newcommand{\cco}{$CC\omega$}
\newcommand{\ccop}{$CC_{\omega}^{+}$}
\newcommand{\shift}[2]{\uparrow_{#2}^{#1}\!}
\newcommand{\shiftnc}[1]{\uparrow^{#1}\!}
\newcommand{\dshiftnc}[1]{\downarrow^{#1}\!}
\newcommand{\nctx}{\Gamma}
\newcommand{\uctx}{\Delta}
\newcommand{\meta}{?}

\begin{document}

\title{Dependently Typed Programming based on Automated Theorem Proving}


\author{Alasdair Armstrong \quad Simon Foster\quad Georg Struth\\
Department of Computer Science, University of Sheffield, UK\\
$\{$a.armstrong,s.foster,g.struth$\}$@dcs.shef.ac.uk
}

\maketitle

\begin{abstract}
\sysname\ is a minimalistic dependently typed programming language and
interactive theorem prover implemented in Haskell. Its main purpose is
to investigate the effective integration of automated theorem provers
in a pure and simple setting. Such integrations are essential for
supporting program development in dependently typed languages. We
integrate the equational theorem prover Waldmeister and test it on
more than $800$ proof goals from the TPTP library. In contrast to
previous approaches, the reconstruction of Waldmeister proofs within
\sysname\ is quite robust and does not generate a significant overhead
to proof search. \sysname\ thus yields a template for integrating more
expressive theorem provers in more sophisticated languages.
\end{abstract}


\section{Introduction}

Dependently typed programming (DTP) languages such as
Adga~\cite{norell09agda} or Epigram~\cite{McBride04} are currently
receiving considerable attention. By combining the elegance of
functional programming with more expressive type systems, they
introduce a new mathematically principled style of program
development. In contrast to traditional functional programming, types
are powerful enough to support detailed specifications of a program's
properties. This however requires type-level reasoning that is no
longer decidable. DTP languages are at the same time interactive
theorem proving (ITP) systems similar to Nuprl~\cite{Nuprl} or
Coq~\cite{bertot2004interactive}. On the one hand this supports
developing programs that are correct by construction. On the other
hand it puts an additional burden on programmers.

To support program development at an appropriate level of abstraction,
it is essential that programmers can focus on the more high-level
creative aspects of proofs, whereas trivial and routine proof tasks
are automated. Yet how can this be achieved?

Traditionally, automation is obtained in ITP systems by implementing large
libraries of tactics, internally verified solvers and sophisticated
simplification techniques, or by using external solvers as
oracles. More recently, external automated theorem proving (ATP)
systems, satisfiability modulo theories (SMT) solvers and other
decision procedures have been integrated in a more trustworthy way into ITP
systems by internally reconstructing proofs provided by the external
tools. A prime example is Isabelle's Sledgehammer tool
(cf.~\cite{BlanchetteBulwahnNipkow11}), which includes a relevance
filter for selecting hypotheses, an interface for passing
proof tasks to external tools, and a mechanism for internally
reconstructing external proofs.

This approach seems particularly promising for DTP languages where it
could make program development more lightweight and less time
consuming. Unfortunately however, ATP integration for DTP languages is
not straightforward. First, state-of-the-art ATP technology is
designed for classical reasoning whereas DTP requires constructive
logic. Second, the logical kernels of DTP languages tend to be much
more complex than those of traditional ITP systems, hence proof
reconstruction establishes relatively less trust. Third, proof
reconstruction turns out to be highly inefficient in practice due to
proof normalisation, whereas in theory it should be linear in the size
of input proofs~\cite{FosterStruth11}.

Due to these issues, ATP integrations for DTP languages certainly
deserve to be studied in a radically pure and simple setting. This
essentially amounts to building a simple trustworthy DTP language
kernel around an ATP system as its most important proof engine. In
this paper we focus in particular on the communication between ATP and
ITP and the efficiency of proof reconstruction. Our main contributions
are as follows.

First, we implement the extended calculus of constructions with
universes as a minimalistic DTP language, called \sysname, in
Haskell. This includes term data types based on de Bruijn indices and
a monadic approach to bidirectional type checking and inference.

Second, we design and implement a simple proof scripting language for
\sysname. It is inspired by Isabelle/Isar and Agda. Apart from
commands for executing interactive proofs it allows calling external
ATP systems within the Proof General interface~\cite{Aspinall00}.

Third, we provide interfaces for executing \sysname\ proofs in the ATP
system Waldmeister and for reconstructing Waldmeister proofs within
\sysname. Proof reconstruction amounts to building a \sysname\ proof
term and type checking it; proof normalisation is avoided.

Fourth, we test the performance of the ATP integration on more than
$800$ proof tasks from the TPTP library~\cite{SS98}. In contrast to
previous approaches, proof reconstruction is very effective and does
not create a significant overhead to Waldmeister proof
search. However, a small number of proof reconstructions currently
fail due to dynamic scoping problems.

In many ways, \sysname\ is still a prototype. The DTP language
implemented has neither recursion nor data types. It is just
expressive enough to support proofs in many-sorted first-order
constructive logic with equality. But for the main purpose of this
paper---exploring effective ATP integrations for DTP lang\-uages---this
is certainly no limitation.

Two particular features of \sysname\ proof reconstruction are that
proof search and proof normalisation are avoided. Our micro-step
reconstruction is in contrast to Isabelle's current macro-step
approach based on the internally verified ATP system Metis~\cite{Hurd05},
and it seems more robust and efficient. In contrast to Agda or
Coq, we only type check the internal proof terms corresponding to
external proofs. These proof terms provide proof certificates that
could be further normalised if needed. Whenever type checking
succeeds, correctness of \sysname's inference system guarantees that
normalisation is possible.


\section{Calculus of Constructions}
\label{Sec:System}

This section introduces the basics of the calculus of construction,
which is \sysname's underlying type theory. We assume familiarity with
basic type systems~\cite{Barendregt:1991,Pierce:2002,Pierce:2005}.
The type inference rules of this calculus are given in Figure
\ref{Fig:cco}; its details are explained in the remainder of this section,

\begin{figure}[ht]

\hrule

\vspace{0.3cm}

\begin{prooftree}
\LeftLabel{\textsc{T-Axiom}}
\RightLabel{$s : s' \in \mathcal{A}$}
\AxiomC{}
\UnaryInfC{$\vdash s \infer s'$}
\end{prooftree}

\vspace{-0.2cm}

\begin{multicols}{2}
\begin{prooftree}
\LeftLabel{\textsc{T-Named}}
\AxiomC{$\named{x} : T \in \nctx$}
\UnaryInfC{$\nctx \vdash \named{x} \infer T$}
\end{prooftree}

\begin{prooftree}
\LeftLabel{\textsc{T-Unnamed}}
\AxiomC{$\uctx \,!\, n \equiv_\beta T$}
\UnaryInfC{$\uctx \vdash n \infer T$}
\end{prooftree}
\end{multicols}

\begin{prooftree}
\LeftLabel{\textsc{$\nctx$-Weakening}}
\RightLabel{$s \in \mathcal{S}$ and $\named{y}$ is fresh}
\AxiomC{$\nctx; \uctx \vdash \named{x} \tycheck T$}
\AxiomC{$\nctx; \uctx \vdash S \infer s$}
\BinaryInfC{$\nctx, \named{y} : S; \uctx \vdash \named{x} \tycheck T$}
\end{prooftree}

\begin{prooftree}
\LeftLabel{\textsc{$\uctx$-Weakening}}
\RightLabel{$\uctx'$ is valid}
\AxiomC{$\uctx \vdash n \infer T$}
\UnaryInfC{$\uctx'\! +\!\!\!\!+ \uctx \vdash n \infer T$}
\end{prooftree}

\begin{prooftree}
\LeftLabel{\textsc{T-Abs}}
\RightLabel{$s \in \mathcal{S}$}
\AxiomC{$\nctx; \uctx \vdash S \infer s$}
\AxiomC{$\nctx; \uctx, S \vdash t \tycheck T$}
\BinaryInfC{$\nctx; \uctx \vdash \lam t \tycheck \piTy S \bdot T$}
\end{prooftree}

\begin{prooftree}
\LeftLabel{\textsc{T-App}}
\AxiomC{$\nctx; \uctx \vdash f \infer \piTy S \bdot T$}
\AxiomC{$\nctx; \uctx \vdash x \tycheck S$}
\BinaryInfC{$\nctx; \uctx \vdash f\,x \infer \;\dshiftnc{1} \subst{0}{\;\shiftnc{1} x}T$}
\end{prooftree}

\begin{prooftree}
\LeftLabel{\textsc{T-Pi}}
\RightLabel{$(s_1, s_2, s_3) \in \mathcal{R}$}
\AxiomC{$\nctx; \uctx \vdash S \infer s_1$}
\AxiomC{$\nctx; \uctx, S \vdash T \infer s_2$}
\BinaryInfC{$\nctx; \uctx \vdash \piTy S \bdot T \infer s_3$}
\end{prooftree}

\begin{prooftree}
\LeftLabel{\textsc{T-Inf}}
\RightLabel{$s \in \mathcal{S}$}
\AxiomC{$\nctx; \uctx \vdash T \infer s$}
\AxiomC{$\nctx; \uctx \vdash t \infer T'$}
\AxiomC{$T \equiv_\beta T'$}
\TrinaryInfC{$\nctx; \uctx \vdash t \tycheck T$}
\end{prooftree}

\begin{prooftree}
\LeftLabel{\textsc{T-Ann}}
\RightLabel{$s \in \mathcal{S}$}
\AxiomC{$\nctx; \uctx \vdash T \infer s$}
\AxiomC{$\nctx; \uctx \vdash t \tycheck T$}
\BinaryInfC{$\nctx; \uctx \vdash t \squaredots T \infer T$}
\end{prooftree}

\hrule

\caption{Typing rules for \cco}
\label{Fig:cco}
\end{figure}

The set of terms of the calculus of constructions ($CC$) is
inductively defined by the following grammar.
\begin{equation*}
 t::= x\in V \; | \; \Pi x:t\ldot t \; | \; \lambda x:t\ldot t \; | \; tt
\end{equation*}
Here, $V$ is a set of variables. $\Pi x:t\ldot t$ is the dependant
product type; it essentially amounts to universal
quantification. $\lambda x:t\ldot t$ is lambda abstraction and $tt$ is
application. In $CC$, types themselves are terms. They are
distinguished, and their mutual dependencies are expressed, by the
type inference rules. Terms that are not types are called
\textit{non-type terms}, or briefly \emph{terms} if the context
allows. A type of a non-type term is called \textit{proper}, whereas
types of types are called \textit{sorts}.

Judgements are expressions $\Gamma \vdash t:T$, where $\Gamma$ is an
environment that provides types for variables, $t$ is a term and $T$ a
type. They can be proved by the type inference rules.

In $CC$, every proper type has sort $\star$, while $\star$ is defined
to have sort $\Box$. The dependencies between terms and types for $CC$
can be modelled by the set
$\{(\star,\star),(\star,\Box),(\Box,\star),(\Box,\Box)\}$. The
statement $(\star,\star)$, for instance, says that terms may depend on
terms; the statement $(\Box,\star)$ says that terms may depend on
types.


To make our calculus rich enough for DTP we extend it with universes.
The \textit{calculus of constructions with universes}, \cco, extends
$CC$ with an infinite set $\Box_0,\dots,\Box_n,\dots$ of
sorts~\cite{Bernardy:2010,Miquel:2001}. A \textit{pure type system}
(PTS) is given by a triple $(\mathcal{S}, \mathcal{A}, \mathcal{R})$,
where $\mathcal{S}$ is a set of sorts and $\mathcal{A}$ a set of
typing relations $s_1 : s_2$ with $s_1,s_2\in\mathcal{S}$. The set
$\mathcal{R}$ consists of triples $(s_1, s_2, s_3)$, where $s_1, s_2,
s_3 \in \mathcal{R}$. This set, in combination with the typing rule
\textsc{T-Pi} in Figure \ref{Fig:cco} controls the dependencies of
terms and types. The PTS for \cco~\cite{Bernardy:2010} is given by:
\begin{align*}
\mathcal{S} &= \{ \star \} \cup \{ \Box_i \mid i \in \mathbb{N} \},\\
\mathcal{A} &= \{ \star : \Box_0 \} \cup \{ \Box_i : \Box_{i+1} \mid i \in \mathbb{N} \},\\
\mathcal{R} &= \{ \star \leadsto \star, \star \leadsto \Box_i,
               \Box_i \leadsto \star \mid i \in \mathbb{N} \} \cup
               \{ (\Box_i, \Box_j, \Box_{\mathrm{max}(i,j)}) \mid i \in \mathbb{N} \},
\end{align*}
The notation $s_1 \leadsto s_2$ is shorthand for
$(s_1,s_2,s_2)$. $\star \leadsto \star$ means that terms can depend on
terms, $\star \leadsto \Box_i$ means that types can depend on terms
(dependent types) and $\Box_i \leadsto \star$ means that terms can
depend on types. The set of all triples $(\Box_i, \Box_j,
\Box_{\max(i,j)})$ defines how types are allowed to depend on
types. If a type $\Box_i$ depends on another type $\Box_j$, it must be
at the same level in the hierarchy as the highest type
$\Box_{\max(i,j)}$ it depends on.

The syntax of \cco\ terms is defined by the following grammar, which
extends and refines that for $CC$:
\begin{align*}
t \Coloneqq\; s \in \mathcal{S}
        \;|\; n \in \mathbb{N}
        \;|\; \named{x} \in V
        \;|\; \lambda t
        \;|\; \piTy t\, .\, t
        \;|\; t\,t
        \;|\; t \squaredots t \ .
\end{align*}
We now use de Bruijn indices to represent variables introduced via
$\lambda$ or $\Pi$ binders, whereas top-level declarations are
named~\cite{Chargueraud:2011}. Named variables are written in
\texttt{typewriter} font; they are elements of $V$, the set of valid
identifiers. Both named and unnamed variables can have free or bound
occurrences. For example, in the term $\lam 3$, the index $3$ is free
because it is pointing outside the term. A term without free de Bruijn
indices is called \emph{locally closed}.

The dependent product type $\piAbs{A}{B}$ corresponds to the logical
statement $\forall a \in A\ldot~B(a)$. To prove $\forall a \in
A\ldot~B(a)$ constructively, one needs to show that for every possible
$a \in A$ an inhabitant of $B$ can be constructed. A function of type
$\piAbs{A}{B}$ is therefore a proof of the statement $\forall a \in A
.~B(a)$. If $B$ does not depend on $A$, then the dependent product
type is $A\rightarrow B$. The additional syntax $t \squaredots t$ is
\emph{type annotation}. It allows us to explicitly state that a given
term has some type.

Because there are two kinds of variables---named and unnamed
ones---judge\-ments take the form $\Gamma; \Delta \vdash t : T$, where
$\Gamma$ and $\Delta$ are the typing contexts for named and unnamed
variables. The syntax for these contexts is
\begin{equation*}
\Gamma \Coloneqq\; \emptyset
             \;|\; \Gamma, \mathrm{x} : T,
\quad\quad
\Delta \Coloneqq\; \emptyset
             \;|\; \Delta, T.
\end{equation*}
Both contexts are lists, but since $\Gamma$ names must be unique in
$\Gamma$, it can be treated as a set. We use $\emptyset$ to represent
empty contexts. We often omit empty contexts and write $\vdash t:T$
rather than $\emptyset;\emptyset\vdash t:T$. We write
$\Gamma,\named{x}:T$ to denote that context $\Gamma$ is extended with
the new binding $\named{x}:T$, whereas for $\Delta$, only a type is
supplied.  We write $\named{x}:T\in\Gamma$ to assert that $T$ is the
type of $\named{x}$ in $\Gamma$. We write $\subst{k}{t'}t$ for the
substitution of term $t'$ for the index $k$ in the term $t$.

Unlike variables in $\Gamma$, those in $\Delta$ are nameless and
cannot be looked up by name. Instead we define two lookup operators
$!$ and $!!$ with and without index shifting to retrieve the types of
variables from $\Delta$.
\begin{equation*}
\Delta, T\, !!\, n = \left\{
  \begin{array}{l l}
    T & \text{if } n = 0,\\
    \Delta\, !!\, (n - 1) & \text{otherwise},\\
  \end{array} \right.
\qquad\qquad
\Delta \, !\, n =\, \shiftnc{n+1}{(\Delta\, !!\, n)}.
\end{equation*}
$\shift{d}{c}t$ is the $d$-place shift of a term $t$ above cutoff
$c$~\cite{Pierce:2002}. We write $\shiftnc{d}t$ if the cutoff is zero.
An unnamed context is well-formed only if all the terms within it are
either proper types or sorts. Unnamed contexts can be concatenated
using the $+\!\!\!+$ operator.


%


To implement this calculus, a \emph{bidirectional type checker} is
used. This means that for any term $t$ of type $T$, one can either
infer the type, written $t \infer T$, or check that the term has the
type, written $t \tycheck T$. Type inference requires $t$ and returns
$T$, whereas type checking requires both $t$ and $T$. The two rules
\textsc{T-Inf} and \textsc{T-Ann} (the rule for type annotations)
provide a conversion between type checking and type
inference. Bidirectional type checking ensures that the rules are
directly implementable without need for further transformation.



\section{An Extended Calculus}

 \sysname\ requires additional features for equational and incremental
 interactive reasoning: identity types and metavariables. We call this
 extension \ccop.

Firstly, we add identity types to \cco. The identity type
$\idTy{A}{a}{b}$ for any type $A$, where $a, b : A$, denotes that $a$
and $b$ represent identical proofs of proposition
$A$~\cite{Nordstrom:1990}. This captures propositional equality within
\sysname\ and supports equational reasoning. Our identity type
corresponds to the implementation of propositional equality as an
inductive family in Agda. Several new terms need to be added to the
grammar of \cco:

\begin{equation*}
t \Coloneqq\; \lam x \ldot t \; | \; \dots \; | \;
           \refl \;
          | \; \idTy{t}{t}{t} \;
          | \; \elimJ      \ .
\end{equation*}

The \textit{reflexivity term} $\refl$ works exactly like {\tt refl} in
Agda~\cite{Tut:Norell:2008}. It allows the construction of identity
types $\idTy{A}{a}{b}$ where $a \equiv_\beta b$. The typing rules for
the reflexivity term \textsc{Eq-Refl} and the identity term
\textsc{Eq-Id} are as follows~\textnormal{\cite{Awodey:2009}}:

\begin{prooftree}
\AxiomC{$\Gamma; \Delta \vdash A \infer s$}
\AxiomC{$\Gamma; \Delta \vdash a, b \tycheck A$}
\AxiomC{$a \equiv_\beta b$}
\LeftLabel{\textsc{Eq-Refl}}
\RightLabel{$s \in \mathcal{S}$}
\TrinaryInfC{$\Gamma; \Delta \vdash \refl \tycheck \idTy{A}{a}{b}$}
\end{prooftree}

\begin{prooftree}
\AxiomC{$\Gamma; \Delta \vdash A \infer s$}
\AxiomC{$\Gamma; \Delta \vdash a, b \tycheck A$}
\LeftLabel{\textsc{Eq-Id}}
\RightLabel{$s \in \mathcal{S}$}
\BinaryInfC{$\Gamma; \Delta \vdash \idTy{A}{a}{b} \infer s$}
\end{prooftree}

The J rule below eliminates identity types~\cite{Dybjer:1994}, which
corresponds to the term $\elimJ$. It can be used in combination with
$\refl$ to define the standard functions of equational logic in
\sysname, namely, substitutivity, congruence, transitivity and
symmetry. Because displaying the J rule with the locally nameless
syntax discussed in Section \ref{Sec:System} would render it almost
unreadable, we present its \sysname\ syntax:


\begin{verbatim}
theorem elimJ : "(A : *) (C : (x y : A) -> Id A x y -> *)
              -> (e : (x : A) -> C x x refl)
              -> (x y : A) (P : Id A x y) -> C x y P".
  "\A C e x y P -> e x".
  qed.
\end{verbatim}

Like Isabelle's proof scripting language Isar, \sysname\ uses two
levels of syntax. The inner syntax, surrounded by quotation marks, is
used for \ccop\ terms. The outer syntax is for proof scripting. For
user interaction, the locally nameless term representation is extended
to a more readable named representation. The inner syntax is
essentially a simplification of Agda's syntax for terms without
implicit arguments or mixfix operators. 

Secondly, metavariables are used in \sysname. Just as in Agda, these
represent ``holes'' within terms that can incrementally be filled
in---or refined---during proofs. Metavariables require one final
language extension:

\begin{equation*}
t \Coloneqq \; \lam x \ldot t \; | \;
            \dots \;
          |\; \meta \ .
\end{equation*}

As an example, consider checking that term $\lam \meta$ has type
$\piAbs{\star}{\piAbs{0}{1}}$.

\begin{prooftree}
\AxiomC{}
\RightLabel{\textsc{T-Axiom}}
\UnaryInfC{$\Gamma; \Delta \vdash \star \infer \Box_1$}
\AxiomC{$\Gamma; \Delta, \star \vdash\; \meta \tycheck \piAbs{0}{1}$}
\RightLabel{\textsc{T-Abs}}
\BinaryInfC{$\Gamma; \Delta \vdash \lam \meta \tycheck \piAbs{\star}{\piAbs{0}{1}}$}
\end{prooftree}
When we try to check that $\meta\! \tycheck\! \piAbs{0}{1}$, the type
checker cannot proceed, so it stores a continuation which allows type
checking to resume once a term for the metavariable has been
supplied. This forms the basis for interactive theorem proving in
\sysname.


\section{Automated Theorem Proving Technology}
\label{Sec:Waldmeister}

Having outlined the type-theoretic foundations of \sysname, we
now discuss the ATP technology which serves as its proof engine.

ATP systems have been designed and implemented for many decades, but
mainly for classical first-order logic with equations. They provide
fully automated proof search based on sophisticated term orderings,
rewriting techniques and heuristics. They can often prove mathematical
statements of moderate difficulty and deal with large hypothesis sets,
which makes them ideally suited for discharging ``trivial''
first-order proof goals in ITP systems. A prime example of an ATP
integration is Isabelle's Sledgehammer tool
(cf.~\cite{BlanchetteBulwahnNipkow11} for an overview), which calls a
number of external ATP systems and SMT solvers. A relevance filter
selects hypotheses for the proof, and the external proof output is
internally reconstructed to increase trustworthiness. Proof
reconstruction is based on the Metis tool~\cite{Hurd05}, an
Isabelle-verified automated theorem prover, which replays the external
proof search with the hypotheses used by the external provers.

An integration of ATP systems into DTP languages is, however, much
less straightforward, as discussed in the introduction. We therefore
start with the simplest case---pure equational logic---for which
classical and constructive reasoning coincide. We integrate the
Waldmeister system~\cite{hillenbrand97waldmeister}, which is highly
effective for this fragment and supports sorts\footnote{We are using
  the last publicly available version of Waldmeister, released in
  1999.}.

Waldmeister accepts a set of equations as hypotheses and a single
equation as a conclusion. It also requires a term ordering to use
rewriting techniques for enhanced proof search. Technically,
Waldmeister is based on the unfailing completion
procedure~\cite{BachmairDershowitzPlaisted89}, a variant of
Knuth-Bendix completion~\cite{KnuthBendix70} that attempts to
construct a (ground) canonical term rewrite system from the equational
hypotheses. This construction need not be finite, but it is guaranteed
that a (rewrite) proof of a valid goal can be found in finite
time. Apart from efficient proof search, Waldmeister offers two
additional features that benefit an integration into \sysname. First,
it provides extremely detailed proof output, down to the level of
positions and substitutions for rewrites in terms. In contrast to
Sledgehammer's macro-step proof reconstruction that replays proof
search, we can therefore check individual proof steps efficiently and
without search. Second, Waldmeister extracts lemmas from proofs. This
memoisation of subproofs further enhances proof reconstruction.

These features can be demonstrated in a simple example from group
theory.  Let $(G, \circ, {}^{-1},1)$ be a group with carrier $G$,
multiplication $\circ$, inversion ${}^{-1}$ and unit $1$. It satisfies
axioms of associativity, right identity and right inverse
\begin{equation*}
  x\circ (y\circ z)=(x\circ y)\circ z,\qquad x\circ 1=x, \qquad x\circ x^{-1}=1.
\end{equation*}
Assume that we have implemented groups in \sysname\ and want to prove
that every right identity is also a left identity: $x^{-1}\circ x =
x\circ x^{-1}$. We then need to pass the axioms and the proof goal to
Waldmeister and let it search for a proof. Figure \ref{Fig:WaldInput}
shows the Waldmeister input file that corresponds to this proof task.

\begin{figure}[thb]
{\footnotesize
\begin{verbatim}
NAME group
MODE PROOF
SORTS
  ANY
SIGNATURE
  e: -> ANY
  i: ANY -> ANY
  f: ANY ANY -> ANY
  a: -> ANY
ORDERING
LPO
  i > f > e > a
VARIABLES
  x,y,z : ANY
EQUATIONS
  f(x,e) = x
  f(x,i(x)) = e
  f(f(x,y),z) = f(x,f(y,z))
CONCLUSION
  f(a,i(a)) = f(i(a),a)
\end{verbatim}
}

\caption{Waldmeister group input file}
\label{Fig:WaldInput}
\end{figure}

The group signature is declared in prefix notation, using sort
\verb|ANY|, and functions \verb|f: ANY ANY -> ANY|,
\verb|i:ANY -> ANY| and \verb|e: -> ANY| for multiplication, inverse,
and unit. An constant \verb|a| is also introduced. Waldmeister's term
ordering is declared in the \verb|ORDERING| block: a lexicographic
path ordering (lpo) is constructed from a precedence on the group
signature and the constant \verb|a|. The next block declares three
variables \verb|x|, \verb|y| and \verb|z| of type \verb|ANY|.  The
\verb|EQUATIONS| block lists the group axioms in Waldmeister syntax.
Finally, the proof goal is declared in Waldmeister syntax for constant
\verb|a|, since universal goals are Skolemised.

After Waldmeister is called, it returns the proof in
Figure~\ref{Fig:WaldOut} within milliseconds.
\begin{figure}[tbh]
{\footnotesize
\begin{verbatim}
  Lemma 1: f(e,i(i(x1))) = x1

    f(e,i(i(x1)))
 =    by Axiom 2 RL at 1 with {x1 <- x1}
    f(f(x1,i(x1)),i(i(x1)))
 =    by Axiom 3 LR at e with {x3 <- i(i(x1)), x2 <- i(x1), x1 <- x1}
    f(x1,f(i(x1),i(i(x1))))
 =    by Axiom 2 LR at 2 with {x1 <- i(x1)}
    f(x1,e)
 =    by Axiom 1 LR at e with {x1 <- x1}
    x1

  Lemma 2: ...

  Lemma 3: ...

  Lemma 4: ...

  Theorem 1: f(a,i(a)) = f(i(a),a)

    f(a,i(a))
 =    by Axiom 2 LR at e with {x1 <- a}
    e
 =    by Axiom 2 RL at e with {x1 <- i(a)}
    f(i(a),i(i(a)))
 =    by Lemma 4 LR at 2 with {x1 <- a}
    f(i(a),a)
\end{verbatim}
}
\caption{Waldmeister group output file}
\label{Fig:WaldOut}
\end{figure}
Here, the \verb|--details| flag has been set to obtain precise
information for each proof step. In the third step of the proof of
\verb|Lemma 1|,
\begin{verbatim}
                   f(x1,f(i(x1),i(i(x1)))) = f(x1,e)
\end{verbatim}
for instance, the right identity axiom \verb|f(x1,i(x1)) = e| has
been used to rewrite from left to right the subterm at position
\verb|2| by matching or substituting $\verb|i(x1)|$ for
\verb|x1|. This level of detail allows efficient micro-step proof
reconstruction; the lemmas generated support proof reconstruction by
memoisation. Details of the communication between \sysname\ and
Waldmeister, in particular proof reconstruction, are covered in the
following section.


\section{Implementing \ccop\ Terms in Haskell}
\label{Sec:HaskellCC}

Users interact with \sysname\ via the Proof General Emacs interface,
which is standard for many ITP systems~\cite{Aspinall00}. User level
terms with explicit variables are parsed to an internal Haskell
representation using de Bruijn indices, as represented by the Haskell
data type \verb|Index|. The complete Haskell implementation can be
found online\footnote{\url{http://www.dcs.shef.ac.uk/~alasdair}}. The
data type has a field \verb|dbInt| for the index and another one,
\verb|dbName|, for the user level variable name. This is useful for
pretty-printing.

{\small
\begin{verbatim}
data Index = DB {dbInt :: Int, dbName :: Text} deriving (Show)

instance Eq Index where
    (DB n _) == (DB m _) = n == m
\end{verbatim}
}

Next we provide data types for sorts and terms.

{\small
\begin{verbatim}
data Sort = Star | Box Word deriving (Show, Eq)

data Term = Sort Sort
          | Unnamed Index
          | Named Text
          | Pi Tag Term Term
          | Ann Term Term
          | App Term Term
          | J Term Term Term Term Term Term
          | Id Term Term Term
          | Lam Tag Term
          | Refl
          | Meta Int
          deriving (Eq, Show)
\end{verbatim}
}
\noindent Tags are used to attach additional information to terms. Specifically,
for \verb|Lam| and \verb|Pi| terms, they store the associated user
level variables, for instance to provide meaningful error
messages. Tags are not relevant for term equality.

Terms can be $\beta$-reduced using the \verb|nf| function. The
function for shifting is implemented as follows:

{\footnotesize
\begin{verbatim}
shift :: Int -> Int -> Term -> Term
shift d c (Unnamed (DB n name))
  | n < c = Unnamed (DB n name)
  | n >= c = Unnamed (DB (n + d) name)
shift d c (Lam tag f)  = Lam tag (shift d (c + 1) f)
shift d c (Pi tag s t) = Pi tag (shift d c s) (shift d (c + 1) t)
shift d c (App f x)    = App (shift d c f) (shift d c x)
shift d c (Ann t ty)   = Ann (shift d c t) (shift d c ty)
shift d c (Id ty a b)  = Id (shift d c ty) (shift d c a) (shift d c b)
shift d c x = x
\end{verbatim}
}


Two additional Haskell functions process metavariables. A first
function generates fresh metavariables as they arise in interactive
proofs. A second function substitutes user supplied expressions for
metavariables. Detailed code can be found at our web site.

The contexts $\Gamma$ and $\Delta$ for named and unnamed variables are
implemented as follows:

{\small
\begin{verbatim}
data Ctx = Ctx { unnamed :: [(Tag, Term)]
               , named   :: OMap Text (Term, Term)
               }

emptyCtx :: Ctx
emptyCtx = Ctx [] OMap.empty
\end{verbatim}
}

Since the order in which variables are added to named contexts may
matter, a custom map data type, \verb|OMap|, has been implemented to
record that information.

Finally, the set $\mathcal{R}$ which defines the dependencies
allowed between types and terms is implemented as follows:

{\small
\begin{verbatim}
setR Star Star       = Star
setR (Box n) Star    = Star
setR Star (Box n)    = Box n
setR (Box n) (Box m) = Box (max n m)
\end{verbatim}
}


\section{Type Checking and Inference in Haskell}

Type checking is performed within a type checking monad. The overall
approach is inspired by that of Agda. The type checking monad
transformer (\verb|TCMT|) is a monad transformer stack consisting of
the \verb|EitherT| monad transformer and the \verb|StateT|
transformer. The state monad carries the type checking context
(\verb|tcmCtx|) as well as lists of inference rules used for type
checking (\verb|tcmTCRules|) and type inference
(\verb|tcmIRules|). It also contains a depth value (\verb|tcmDepth|)
for tracing and logging the type checking process
(\verb|tcmLog|). Whenever a metavariable is encountered during
type checking, a continuation is added to the state (in
\verb|tcmMetas|). It contains the information required to resume type
checking once a user supplies a value for it. As mentioned in Section
\ref{Sec:HaskellCC}, metavariables must be fresh, so a counter is used
for indexing them. The either monad allows handling failures; when
type checking fails we use it to return \verb|TypeError| values.

{\small
\begin{verbatim}
newtype TCMT m a = TCMT
    { unTCMT :: EitherT TypeError (StateT (TCMState m) m) a
    } deriving (Functor, Applicative, Monad, MonadIO)

data TCMState m = TCMState { tcmDepth   :: Int
                           , tcmCtx     :: Ctx
                           , tcmTCRules :: [TCRule m]
                           , tcmIRules  :: [IRule m]
                           , tcmMetas   :: [MetaContinuation]
                           , tcmLog     :: [LogEntry]
                           , tcmCounter :: Counter
                           }

data MetaContinuation = MC Ctx Int Term
\end{verbatim}
}

Type checking rules have the form \verb|Term -> Term -> TCMT m Bool|,
where, as mentioned in Section \ref{Sec:System}, both the term and its
tentative type are provided as inputs. Type checking returns
\verb|True| if the terms type check, and \verb|False| if rule
application fails (in which case another rule will be selected). It
fails with \verb|TypeError| if a term does not type check. Type
inference rules require only a term $t$ as an input. They return
\verb|Just| $T$ when $t \infer T$, and \verb|Nothing| when the
inference rule cannot be applied; \verb|TypeError| is raised when
the rule fails.

{\small
\begin{verbatim}
data TCRule m = TCR { ruleName :: Text
                    , rule :: Term -> Term -> TCMT m Bool
                    }
\end{verbatim}

\begin{verbatim}
data IRule m = IR { inferRuleName :: Text
                  , inferRule :: Term -> TCMT m (Maybe Term)
                  }
\end{verbatim}
}

Two Haskell functions are used for type checking and type inference
with the \verb|TCMT| monad. The \verb|typecheck| function takes two
terms as arguments and attempts to apply a type checking inference
rule. It returns \verb|Nothing| if no inference rule can be found, and
the name of the rule otherwise. The \verb|hasType| function is
similar, but simply fails if no rule can be applied. It is usually
called as an infix function, and allows rules to be written in a more
declarative fashion.

{\small
\begin{verbatim}
typecheck :: (Functor m, Monad m) => Term -> Term -> TCMT m (Maybe Text)
typecheck t1 t2 = do
    tRules <- tcmTCRules <$> get
    foldM tryRule Nothing tRules
  where
    tryRule (Just name) _ = return (Just name)
    tryRule Nothing (TCR name rule) = do
        r <- rule t1 t2
        return $ if r then Just name else Nothing
\end{verbatim}
}

The \verb|infer| function attempts to infer the type of its
argument. It fails if no inference rule can be applied and returns the
inferred term otherwise.

{\small
\begin{verbatim}
infer :: (Functor m, Monad m) => Term -> TCMT m Term
infer t | inf t = do
    iRules <- tcmIRules <$> get
    r <- foldM tryRule Nothing iRules
    case r of
      (Just t) -> return t
      Nothing -> __ERROR__ "infer" [("t", t)]
                 "no rule could be applied to infer the type of\n{t}"
  where tryRule (Just t) _ = return (Just t)
        tryRule Nothing (IR name rule) = rule t
\end{verbatim}
}

As an example of a type checking rule, the code for the \textsc{T-Abs}
rule from Figure \ref{Fig:cco} is shown below. Pattern matching and
guards are used to restrict the terms it can be applied to. Each line
in the do block then imposes such a condition. \verb|validType| checks
that \verb|argType| is either a proper type or a sort, while the next
line checks that the body of the lambda expression has the correct
type. If both these conditions hold, the rule can be applied and
\verb|True| is returned.

{\small
\begin{verbatim}
tAbs ::  (Functor m, Monad m) => Term -> Term -> TCMT m Bool
tAbs (Lam tag expr) pi@(Pi _ argType exprType) | inf pi = do
    validType argType
    withUnnamedVar tag argType $ expr `hasType` exprType
    return True

tAbs _ _ = return False

tAbsRule :: (Functor m, Monad m) => TCRule m
tAbsRule = TCR "T-Abs" tAbs
\end{verbatim}
}






This Haskell infrastructure suffices to implement the \ccop part of
\sysname. The ATP integration is described in the next section.


\section{ATP Integration}
\label{Sec:ATPIntegration}

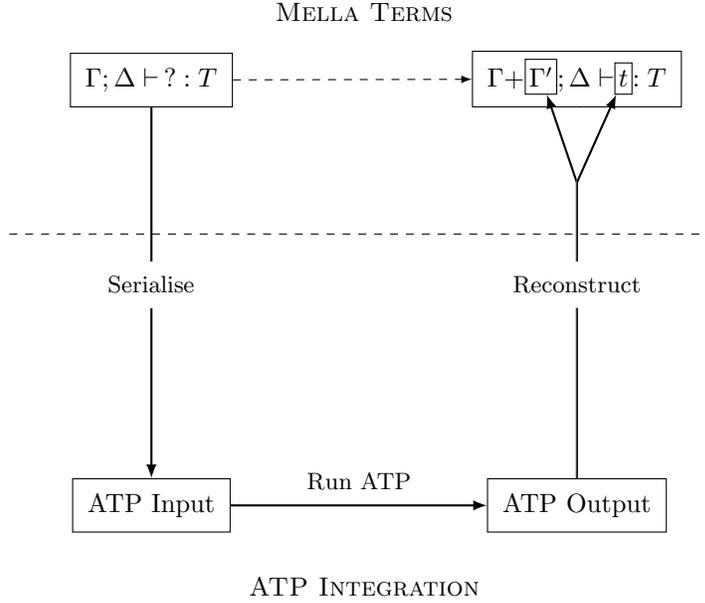
\begin{figure}[bthp]
\centering
\begin{tikzpicture}[inner sep=2mm,baseline,anchor=base,>=latex]
  \node [rectangle, draw] (nodeGoal) at (135:4cm) {$\Gamma;\Delta \vdash\, ? : T$};

  \node [rectangle, draw] (nodeATPI) at (-135:4cm) {ATP Input};

  \matrix [inner ysep=1mm, rectangle, draw] (matProof) at (45:4cm)
  {
    \node [inner xsep=0mm, inner ysep=1mm] {$\Gamma+$}; &
    \node [inner xsep=0.5mm, inner ysep=1mm, rectangle, draw] (nodeGamma) {$\Gamma'$}; &
    \node [inner xsep=0mm, inner ysep=1mm] {$; \Delta \vdash$}; &
    \node [inner xsep=0.5mm, inner ysep=1mm, rectangle, draw] (nodet) {$t$}; &
    \node [inner xsep=0mm, inner ysep=1mm] {$: T$}; \\
  };

  \node [below=1cm of matProof, minimum height=0mm] (lineSplit) {};

  \node [rectangle, draw] (nodeATPO) at (-45:4cm) {ATP Output};

  \draw[->, thick] (nodeATPI) -- (nodeATPO) node [midway, above] {\small{Run ATP}};

  \draw[thick] (nodeATPO) -- (lineSplit.north);
  \draw[->, thick] (lineSplit.north) -- (nodet);
  \draw[->, thick] (lineSplit.north) -- (nodeGamma);
  \draw[opacity=0] (nodeATPO) -- (matProof) node [midway, opacity=1, fill=white] {\small{Reconstruct}};

  \draw[->, thick] (nodeGoal) -- (nodeATPI) node [midway, fill=white] {\small{Serialise}};

  \draw[->, dashed] (nodeGoal) -- (matProof);

  \node[anchor=south] (nodeTerm) at (90:3.5cm) {\textsc{\sysname\ Terms}};

  \node[anchor=north] (nodeTactic) at (-90:3.5cm) {\textsc{ATP Integration}};

  \node (nodeLeftDash) at (170:5cm) {};
  \node (nodeRightDash) at (10:5cm) {};
  \draw[dashed] (nodeLeftDash) -- (nodeRightDash);
\end{tikzpicture}
\caption{Overview of Waldmeister Integration}
\label{Fig:Overview}
\end{figure}

\noindent Our general approach to ATP integration is depicted in
Figure~\ref{Fig:Overview}.  \sysname\ proof tasks are represented as
judgements $\Gamma;\Delta\vdash \, ?:T$. They encode that from a set
of hypotheses given by the contexts $\Gamma$ and $\Delta$ a proof term
$t$---represented by metavariable $?$---of type $T$ (the proof goal)
is to be inferred. This is achieved by \emph{serialising} $\Gamma$,
$\Delta$ and $T$ and passing them on to Waldmeister. In our group
example, $\Gamma$ and $\Delta$ contain the group axioms, whereas $T$
contains the proof goal. More generally, the contexts can also contain
lemmas that have been proved before. If Waldmeister fails to find a
proof within a certain time limit, the user is notified. Otherwise,
its proof output is translated into a proof term $t$ in \sysname,
which is then type checked. Since Waldmeister produces intermediate
lemmas, as we have seen, an additional context $\Gamma'$ is added to
$\Gamma$. Constructing a proof term from a Waldmeister proof and type
checking it yields \emph{proof reconstruction}.  We now discuss the
individual steps in more detail.

A \sysname\ file consists of a list of commands delimited by periods,
each of which can be processed and undone individually by Proof
General. There are about 20 commands available to the user, which can
be displayed using the \verb|commands| command. The \verb|help|
command provides a documentation for every command in the system. The
command \verb|fun| introduces a new top-level function or value. The
following commands, for instance, introduce an identity function and a
constant function in \sysname.  {\small
\begin{verbatim}
fun id : "(A : *) -> A -> A"
  "\_ x -> x".

fun const : "(A B : *) -> A -> B -> A"
  "\_ _ x _ -> x".
\end{verbatim}
}To declare a theorem and start a proof, the \texttt{theorem} command
is used. It takes the name of the theorem and its type $T$. To prove
the theorem, the user must construct a proof term $t$ such that $t
\tycheck T$. Proofs are built up incrementally from commands and terms
that may themselves contain metavariables. 

As an example, assume we
want to prove that
\begin{equation*}
f(x,g(y,g(x,z)))= x\qquad\text{ and }\qquad
g(x,f(y,f(x,z)))=x
\end{equation*}
imply
\begin{equation*}
f(x,g(y,x))=x.
\end{equation*}
A ``manual'' \sysname\ proof without using Waldmeister is as follows:
{\small
\begin{verbatim}
theorem example : "(A : *) (f g : A -> A -> A)
                -> (axiom1 : (x y z : A) -> Id A (f x (g y (g x z))) x)
                -> (axiom2 : (x y z : A) -> Id A (g x (f y (f x z))) x)
                -> (x y t1 t2 : A) -> Id A (f x (g y x)) x".
  intro A f g ax1 ax2 x y t1 t2.
  = "f x (g y (g x (f t1 (f x t2))))" by "ax2 x t1 t2" at '2,2RL'.
  = "x" by "ax1 x y (f t1 (f x t2))".
  refl.
  qed.

normalize proof.
describe proof.
\end{verbatim}
}
\noindent \sysname\ commands can be terms, which are surrounded by quotation
marks, theorem definitions, function definitions, or command
expressions.
The command \verb|intro| $args$ generates a term of the
form $\lam\mathit{args}\rightarrow\; ?$. The command {\small
\begin{verbatim}
 = "f x (g y (g x (f t1 (f x t2))))" by "ax2 x t1 t2" at '2,2RL'.
\end{verbatim}
}
\noindent says that the left-hand term in the proof goal is equal to
the term provided by applying the second axiom at position \verb|2,2|
to variable \verb|x| from right to left, in a notation similar to
Waldmeister. The second proof step is similar. The remaining step is
reflexivity of equality. The commands \verb|normalize proof| and
\verb|describe proof| normalise the proof and print out the proof term
(which we do not show). We can also use the \verb|agda| command to
compile \sysname\ files into Agda files. This is very useful for
testing the correctness of our implementation.

Command expressions form a large part of \sysname's syntax. Examples
are \verb|intro|, \verb|=| and \verb|qed| as displayed above. Commands
consist of a command name, followed by zero or more arguments and a
list of keywords. Each keyword can again be associated with a list of
arguments:

\begin{equation*}
\mathtt{command} \; arg_1 ... arg_n \; \mathtt{:\!keyword1} \; karg_1 ... karg_n \; \mathtt{:\!keyword2} \; ...
\end{equation*}

Alternatively to the above manual proof we can use Waldmeister to
prove our goal.

{\small
\begin{verbatim}
theorem example : "...".
  intro A f g ax1 ax2 x y.
  waldmeister :signature f g x y :axioms ax1 ax2 :kbo :timeout 2.
  qed.
\end{verbatim}
}

\noindent The \verb|waldmeister| command is now used to instantiate
the metavariable opened by the \verb|intro| command. Waldmeister is
given the functions and values it may use in the proof via the
\verb|:signature| keyword, which maps to the \verb|SIGNATURE| section
of the Waldmeister input file. The axioms to be used when constructing
the proof are listed after the \verb|:axioms| keyword, and are used in
the \verb|EQUATIONS| section of the Waldmeister input file. The
\verb|:kbo| option tells Waldmeister to use a Knuth-Bendix ordering as
the syntactic ordering for terms (based on the precedence given by the
order of expressions declared after \verb|:signature|). Finally, the
\verb|:timeout| keyword lets one specify the amount of time
Waldmeister will be given for proof search.

We now describe proof reconstruction. As already mentioned,
Waldmeister splits proofs into lemmas. While this process is primarily
intended to increase readability, it also enhances proof
reconstruction by memoising subproofs.

\begin{sloppypar}
The Waldmeister output for the example proof above is
shown below. Waldmeister renames variables in its output, so during
reconstruction, the renamed variables must be matched with the correct
variables within \sysname. The proof shows that the term
\verb|s5(s1,s4(s0,s1)| is equal to \verb|s1|. It consists of
two steps: First, Waldmeister applies Axiom 2 from right to left at position
\verb|2.2| in \verb|s5(s1,s4(s0,s1)|, which results in the term shown
on the next line. Secondly, Waldmeister uses Axiom 1 to reduce the term down to
\verb|s1|, proving the goal.
\end{sloppypar}

{\small
\begin{verbatim}
  Theorem 1: s5(s1,s4(s0,s1)) = s1

    s5(s1,s4(s0,s1))
 =    by Axiom 2 RL at 2.2 with {x3 <- y, x2 <- z, x1 <- s1}
    s5(s1,s4(s0,s4(s1,s5(z,s5(s1,y)))))
 =    by Axiom 1 LR at e with {x3 <- s5(z,s5(s1,y)), x2 <- s0, x1 <- s1}
    s1
\end{verbatim}
}

To prove a goal $x = y$, each step of a proof applies a lemma or axiom
to a subterm of $x$. In \sysname\ this requires us to use the
inference rules for congruence (to select the subterm) and symmetry
(to choose the direction). If neither congruence nor symmetry is
required for a step, they are omitted from the proof output, as is
the case for the second step above. The above Waldmeister proof has
two steps, hence we need to use transitivity to join both steps
together, resulting in the final reconstructed \sysname\ proof term
below. This proof term is somewhat unreadable; it has been indented to
make the structure of the proof clearer.

{\small
\begin{verbatim}
trans A (f x (g y x)) (f x (g y (g x (f y (f x y))))) x
  (cong A A x (g x (f y (f x y))) (\rc-cong-var -> f x (g y rc-cong-var))
    (sym A (g x (f y (f x y))) x
      (ax2 x y y)))
  (ax1 x y (f y (f x y)))
\end{verbatim}
}




\section{Proof Experiments}
\label{Sec:Results}

We tested the Waldmeister integration on $850$ proof goals from the
TPTP library~\cite{SS98}, among them $115$ on Boolean algebras (BOO),
$156$ on lattices (LAT), $415$ on groups (GRP), $106$ on relation
algebras (REL) and $58$ on rings (RNG). The letters in brackets
indicate the name given to these problem sets in TPTP. The library
contains non-theorems and non-equational theorems that are beyond
Waldmeister's scope. In fact, in our experiments, Waldmeister has not
been able to find proofs for all goals for principal reasons, but may
also have failed to find proofs of equational theorems due to
timeout. Here, however, we are only interested in relative success
rates for proof reconstruction, that is, the number or percentage of
successful Waldmeister proofs that \sysname\ was able to reconstruct,
and in the running times of proof reconstruction relative to proof
search. The outcome of these experiments are shown in
Table~\ref{Tbl:Res}.
\begin{table}[htb]
\footnotesize
\centering
\begin{tabular}{ l || l | l | l | l || l | l | l }
&\multicolumn{4}{|c||}{Waldmeister}&\multicolumn{3}{c}{Proof Reconstruction}\\
\hline
       & CPU Time & Timeout & Error & Unprovable & Fail & Success &  \% \\
\hline\hline
BOO    & 300     & 9        & 51     & 1           & 10      & 44       & 81.5  \\
   & 30      & 41       & 21     & 1           & 10      & 42       & 80.8  \\
   & 10      & 62       & 0      & 1           & 10      & 42       & 80.8  \\
   & 5       & 64       & 0      & 1           & 10      & 40       & 80    \\
    & 1       & 66       & 0      & 1           & 10      & 38       & 79.2  \\
\hline
LAT    & 300     & 91       & 14     & 0           & 11      & 40       & 78.4  \\
    & 30      & 107      & 0      & 0           & 11      & 38       & 77.6  \\
    & 10      & 110      & 0      & 0           & 11      & 35       & 76.1  \\
    & 5       & 110      & 0      & 0           & 11      & 35       & 76.1  \\
   & 1       & 113      & 0      & 0           & 9       & 34       & 79.1  \\
\hline
GRP    & 300     & 39       & 4      & 0           & 213     & 159      & 42.7  \\
    & 30      & 53       & 1      & 0           & 202     & 159      & 44.0  \\
    & 10      & 57       & 0      & 0           & 199     & 159      & 44.4  \\
    & 5       & 66       & 0      & 0           & 192     & 157      & 45    \\
   & 1       & 91       & 0      & 0           & 185     & 139      & 42.9  \\
\hline
REL    & 300     & 20       & 2      & 0           & 2       & 82       & 97.6  \\
    & 30      & 34       & 0      & 0           & 2       & 70       & 97.2  \\
    & 10      & 58       & 0      & 0           & 2       & 46       & 95.8  \\
    & 5       & 61       & 0      & 0           & 0       & 45       & 100   \\
    & 1       & 66       & 0      & 0           & 0       & 40       & 100   \\
\hline
RNG    & 300     & 35       & 5      & 0           & 0       & 18       & 100   \\
    & 30      & 41       & 0      & 0           & 0       & 17       & 100   \\
    & 10      & 44       & 0      & 0           & 0       & 14       & 100   \\
    & 5       & 44       & 0      & 0           & 0       & 14       & 100   \\
    & 1       & 45       & 0      & 0           & 0       & 13       & 100   \\
\end{tabular}
\caption{Proof Reconstruction Experiments}
\label{Tbl:Res}
\end{table}

The first column in the table shows the TPTP problem sets. The first
four columns are related to Waldmeister. The first of them shows the
Waldmeister CPU time limits for proof search---$1s$, $5s$, $10s$,
$30s$ and $300s$. The second one gives the number of proofs searches
that exceeded the time limit. The third one gives the number of proofs
that aborted, for instance, due to out of memory errors. In the case
of Boolean algebras, the fourth row shows that Waldmeister refuted one
proof goal. The final three columns contain data on proof
reconstruction. The first of them shows the number of proofs for which
reconstruction failed; the second one the number of successfully
reconstructed proofs. The row-wise sums of these columns give the
numbers of successful Waldmeister proofs. The third row gives the
percentage of successful proof reconstructions.

First, it turns out that the CPU time limit for Waldmeister has little
impact on success rates. The number of successful Waldmeister proofs
increases only slightly with proof search time; the success rates for
reconstruction remain almost unaffected. This suggests that there is
little correlation between proof search time and the difficulty of
reconstructing the resulting proof. Waldmeister could spend a long
time traversing a search space only to find a very short and simple
proof which is trivial to reconstruct.

Second, success rates are surprisingly different for different problem
sets. For groups, proof reconstruction was particularly poor,
succeeding only 45\% of the time for proofs returned after a 5 second
timeout. For rings and relation algebras, reconstruction succeeded
almost always, with a 100\% reconstruction success rate at 5
seconds. For lattices and Boolean algebras reconstruction was also
overall successful; it is 80\% for Boolean algebras and 76.1\% for
lattices (again with a 5 second timeout). Some explanations for this
are given below.

\begin{figure}[htb]
\begin{tikzpicture}[x=7mm, y=7mm, scale=2.5, >=stealth]
\draw (0,0) -- (5, 0);
\draw (0,0) -- (0, 5);
\foreach \x in {0,1,...,5} \draw (\x,1pt) -- (\x,-3pt) node[anchor=north] {\x};
\foreach \y in {0,1,...,5} \draw (1pt,\y) -- (-3pt,\y) node[anchor=east] {\y};
\node[below=0.8cm] at (2.5,0) {Waldmeister Running Time (s)};
\node[rotate around={90:(0,0)}] at (-0.4cm, 2.5) {Reconstruction Time (s)};
\path[mark=x, mark size=1] plot coordinates {
(0.005871,0.004081)
(0.014778,0.043859)
(4.834003,0.684167)
(0.029679,0.145771)
(0.010846,0.040701)
(0.000304,0.003794)
(0.000692,0.001071)
(0.025714,0.184832)
(0.023807,0.154592)
(0.003866,0.054217)
(0.000723,0.012955)
(0.013909,0.036159)
(0.025001,0.146287)
(0.003979,0.049401)
(0.01068,0.049973)
(0.236687,0.588522)
(0.045179,0.06849)
(1.972507,0.627011)
(0.103377,0.754087)
(0.019184,0.087156)
(0.064476,0.288852)
(2.65433,1.637759)
(0.11865,0.303243)
(0.067897,0.130446)
(0.244361,0.834965)
(0.122174,0.455835)
(0.219649,0.489409)
(0.971681,1.289491)
(0.000822,0.001494)
(0.019292,0.100643)
(0.297219,0.316335)
(0.050503,0.173808)
(0.07153,0.374677)
(0.116556,0.356312)
(0.065996,0.387365)
(0.347326,0.971447)
(0.960711,2.092128)
(0.039362,0.149227)
(0.351763,0.968545)
(1.170436,0.90042)
(2.086082,1.046559)
(0.070033,0.263634)
(2.018039,0.918238)
(0.613649,1.30454)
(0.580565,1.216273)
(0.007022,0.010502)
(0.009691,0.006795)
(0.009906,0.054456)
(0.6052,1.633133)
(0.116484,0.585377)
(0.568167,1.163758)
(0.117749,0.605458)
(0.076591,0.42188)
(0.051152,0.402202)
(0.190635,0.363355)
(0.202756,0.371607)
(0.130161,0.492359)
(0.129813,0.622876)
(0.130497,0.67232)
(0.014141,0.132828)
(0.012011,0.143149)
(1.692169,0.332998)
(0.005138,0.022061)
(0.033585,0.319821)
(0.003251,0.001285)
(0.004961,0.001139)
(0.006679,0.001976)
(0.767889,1.027025)
(0.00361,0.000834)
(0.009351,0.016373)
(0.004723,0.007381)
(0.006152,0.001222)
(0.005483,0.004593)
(0.235906,0.68472)
(0.013661,0.123043)
(0.004137,0.00108)
(0.003239,0.001298)
(0.004288,0.0038)
(0.004056,0.001217)
(0.759361,0.973362)
(0.235984,0.655809)
(0.015326,0.081183)
(0.004116,0.00273)
(0.003166,0.001754)
(0.017914,0.110979)
(0.003615,0.001145)
(0.01227,0.010115)
(1.697489,0.311645)
(0.004446,0.004255)
(0.004322,0.001672)
(0.231685,0.48683)
(0.010352,0.008935)
(0.00438,0.001057)
(0.003651,0.001369)
(0.003673,0.015465)
(0.004702,0.007439)
(0.016283,0.147446)
(0.006589,0.061207)
(0.826325,0.189223)
(0.004856,0.001426)
(0.004512,0.001555)
(0.118174,0.775094)
(0.004617,0.001956)
(0.004304,0.002134)
(0.007671,0.061712)
(0.005037,0.002824)
(0.004532,0.003137)
(0.850174,0.190844)
(0.004577,0.002898)
(0.01195,0.011159)
(0.172293,0.15331)
(0.004865,0.002064)
(1.678571,0.273829)
(0.005778,0.006366)
(1.836033,0.347033)
(1.687628,0.329021)
(0.004584,0.00152)
(1.679931,0.349307)
(0.030927,0.100064)
(0.000702,0.002391)
(0.009234,0.04338)
(0.163254,0.157637)
(0.005973,0.004901)
(0.159814,0.158599)
(0.003778,0.007789)
(0.008669,0.005406)
(1.706924,0.583532)
(0.004047,0.001683)
(1.706713,0.538629)
(1.744541,0.36146)
(0.028905,0.090054)
(0.785902,0.14663)
(0.78093,0.139742)
(0.044784,0.123147)
(3.698517,0.434138)
(0.004617,0.004216)
(2.22826,0.115292)
(0.165345,0.158483)
(0.004089,0.001751)
(0.004044,0.001937)
(0.004545,0.001897)
(0.1655,0.159741)
(0.028512,0.091334)
(1.831764,0.326404)
(0.16378,0.235202)
(1.138256,0.330831)
(0.02456,0.089813)
(1.143383,0.35623)
(0.011326,0.043039)
(0.004893,0.001906)
(1.692519,0.4545)
(0.006578,0.020578)
(1.147708,0.374294)
(0.002016,0.004974)
(3.725669,0.569854)
(1.138665,0.369234)
(0.006401,0.013338)
(0.005348,0.039066)
(0.001065,0.00685)
(0.005553,0.023851)
(0.004925,0.013711)
(0.004954,0.023686)
(0.005504,0.034889)
(0.016132,0.04629)
(0.004232,0.001677)
(0.001763,0.022395)
(0.000795,0.010007)
(0.001924,0.032951)
(0.00336,0.005228)
(0.002584,0.046639)
(0.004231,0.015217)
(0.006098,0.031657)
(0.00625,0.019543)
(0.014316,0.046146)
(0.007117,0.035456)
(0.003604,0.001824)
(0.004743,0.013221)
(0.001912,0.026783)
(0.004786,0.013217)
(0.001883,0.026547)
(0.019618,0.018482)
(0.003605,0.001852)
(0.00573,0.00848)
(0.003438,0.00198)
(0.005494,0.001877)
(0.00212,0.031844)
(0.004627,0.022185)
(0.003897,0.0071)
(0.003754,0.005115)
(0.00332,0.003128)
(0.006706,0.02933)
(0.003438,0.003161)
(0.005026,0.020575)
(0.003581,0.002935)
(0.009261,0.033265)
(0.010062,0.053283)
(0.024278,0.056884)
(0.004909,0.018516)
(0.014418,0.076308)
(0.021338,0.040248)
(0.008451,0.033053)
(0.004454,0.016335)
(0.017628,0.045398)
(0.002358,0.029738)
(0.012028,0.043671)
(0.00516,0.030781)
(0.004034,0.004984)
(0.010919,0.067143)
(0.0066,0.040784)
(0.009175,0.039521)
(0.00103,0.005777)
(0.00148,0.016312)
(0.002214,0.036007)
(0.004048,0.002426)
(0.002573,0.019926)
(0.001857,0.022499)
(0.001516,0.017378)
(0.193413,2.777872)
(0.02112,0.098629)
(0.185081,2.698772)
(0.221377,4.10272)
(0.014948,0.101536)
(0.005674,0.014648)
(0.19184,2.751827)
(0.193865,2.776487)
(2.126015,0.173236)
(0.096043,0.411702)
(0.003175,0.001027)
(0.008061,0.037793)
(0.037733,0.081026)
(0.003437,0.00576)
(0.364555,0.076037)
(0.273414,0.092936)
(0.004855,0.000964)
(0.408747,0.099115)
(0.004262,0.017552)
(0.303433,0.053285)
(0.003276,0.001236)
(0.003566,0.000866)
(0.012234,0.028538)
(0.365694,0.257008)
(0.004336,0.001741)
(0.077342,0.099159)
(0.031602,0.110332)
(0.523097,0.152114)
(0.005526,0.003512)
(0.024001,0.01738)
(0.620292,0.17509)
(0.008892,0.011001)
(0.038303,0.045983)
(0.361134,0.243877)
(0.009715,0.040518)
(0.00636,0.017033)
(0.005872,0.024458)
(0.153222,0.20524)
(0.057534,0.185961)
(0.008195,0.003115)
(0.005198,0.001355)
(2.672615,0.287751)
(0.005065,0.016153)
(0.392162,0.142706)
(0.008555,0.036759)
(0.699807,0.268482)
(0.003236,0.018364)
(4.293913,0.600245)
(0.183921,0.248439)
(0.004192,0.006931)
(0.00404,0.004997)
(0.005719,0.006549)
(0.020617,0.118871)
(0.005448,0.016758)
(0.010633,0.026274)
(0.004142,0.044906)
(0.695711,0.240693)
(0.01114,0.030676)
(0.008796,0.032989)
(0.105696,0.074523)
(0.021518,0.178915)
(0.009014,0.03223)
(0.022208,0.121849)
(0.005912,0.031672)
(0.02331,0.104087)
(0.004856,0.023969)
(0.04428,0.153189)
(0.055376,0.34176)
(0.003254,0.005582)
(0.011344,0.010287)
(0.009939,0.041869)
(0.103331,0.056047)
(0.022183,0.102099)
(0.006773,0.011066)
};
\end{tikzpicture}
\caption{Waldmeister running times versus proof reconstruction times}
\label{Fig:Graph}
\end{figure}
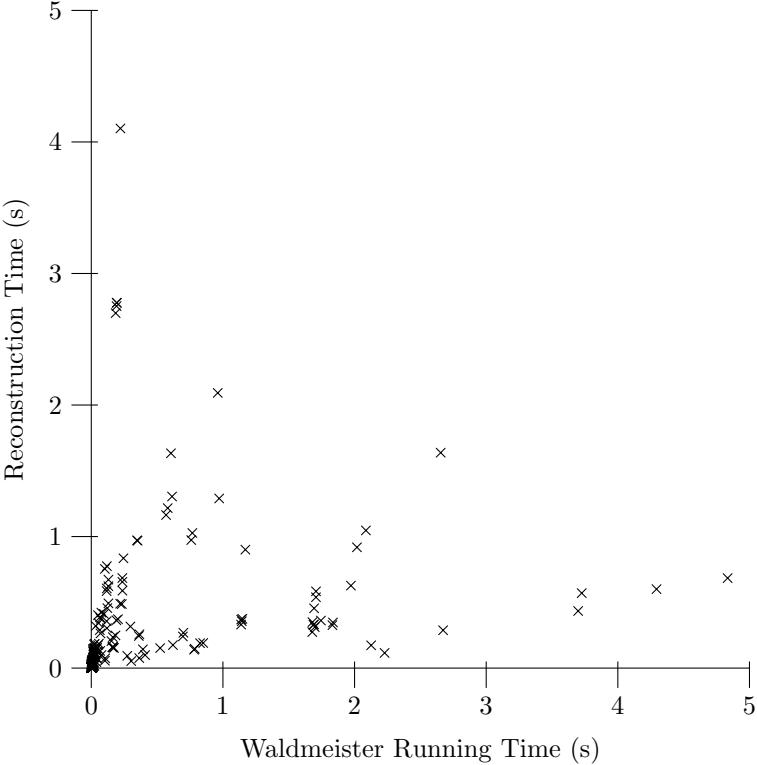

Next we have investigated the correlation between proof search and
proof reconstruction times. A graph is plotted in
Figure~\ref{Fig:Graph}. Unfortunately, these times were very short for
most of our proofs, which makes it very difficult to draw convincing
conclusions.  For some proofs, proof search took rather long whereas
reconstruction was fast. In other cases, proof search was fast, but
the proof could not be reconstructed or type checked efficiently. We
have inspected the proof for each goal that took longer than $2s$ to
reconstruct. In each of these cases, either proof terms are extremely
long, with more than $100$ lemmas, or there are extremely large
substitutions.

As an example, consider the following line from Figure
\ref{Fig:WaldOut}:

\begin{verbatim}
 =  by Axiom 2 LR at 2 with {x1 <- i(x1)}
\end{verbatim}
\noindent In the substitution \verb|x1 <- i(x1)|, for instance, the
term \verb|i(x1)| can be enormous. In fact, our experiments contain
substitutions of terms thousands of characters long, resulting in
extremely large and unwieldy lemmas. This underscores the benefit of
Waldmeister's lemma generation, which allows us to type check each one
individually. As soon as proof terms become large, type checking slows
down. These observations confirm what one would expect: proof
reconstruction times depend on proof sizes rather than proof search
times, whereas proof search time and proof size are often only weakly
correlated. Proof length, however, is not a key factor for using ATP
systems in DTP program development. Ultimately, our experiments
suggest that the Waldmeister integration into \sysname\ is feasible,
and proof reconstruction yields little overhead to proof search.

There are several reasons why proof reconstruction may fail. Firstly,
Waldmeister sometimes introduces fresh Skolem constants in
proofs. These currently cannot be handled by the proof reconstruction
code and cause it to fail. More precisely, such constants, which are
dynamically generated by Waldmeister can currently not be associated
with an environment during proof reconstruction. Secondly, rules such
as the right inverse axiom $x\circ x^{-1}=1$ for groups, when applied
from right to left to a (sub)term $1$, can lead to ``inventing'' fresh
variables $x$ in a Waldmeister proof. \sysname\ would then have to
introduce this value to the type signature of the lemma and supply it
as a parameter. This currently assigns lemmas the wrong types in
proofs and causes proof reconstruction to fail. For certain problem
sets such as groups, ``creative'' proof steps of this kind seem
particularly frequent, whereas in others (such as Boolean algebras,
relation algebras or rings), they are present, but seem less
significant.

As an example, consider the proof discussed in Section \ref{Sec:ATPIntegration}:

{\small
\begin{verbatim}
theorem proof : "(A : *) (f g : A -> A -> A)
              -> (axiom1 : (x y z : A) -> Id A (f x (g y (g x z))) x)
              -> (axiom2 : (x y z : A) -> Id A (g x (f y (f x z))) x)
              -> (x y : A) -> Id A (f x (g y x)) x".
  intro A f g ax1 ax2 x y.
  waldmeister :signature f g x y :axioms ax1 ax2 :kbo :timeout 2.
  qed.
\end{verbatim}
}

\noindent Waldmeister uses the following lemma in its proof:

{\small
\begin{verbatim}
  Lemma 1: s10(x1,s14(x2,x1)) = x1

    s10(x1,s14(x2,x1))
 =    by Axiom 7 RL
    s10(x1,s14(x2,s14(x1,s10(z,s10(x1,y)))))
 =    by Axiom 8 LR
    x1
\end{verbatim}
}

\noindent The second line of this proof introduces the new variables
$y$ and $z$. They are not mentioned in that lemma's type, hence the
lemma cannot be easily reconstructed. We have implemented heuristics
that guess instances of correct type for $z$ and $y$ (in this case
$x1$ and $x2$) which are present in the context. In this particular
lemma, these heuristics make proof reconstruction succeed. In many
other case, we still obtain confusing error messages.


\section{Related Work}

The general question of proof automation for ITPs is covered in a wide
variety of literature. Barendregt and Barendsen \cite{Barendregt02}
identify three approaches, namely \emph{accepting}, \emph{skeptical},
and \emph{autarkic}. The accepting approach uses ATPs and SMT solvers
as oracles, requiring no proof output. The skeptical approach requires
that external tools provide evidence or certificates which allow ITP
systems to internally reconstruct external proofs to increase
trust. The autarkic approach solely relies on internal implementations
of solvers and provers or alternatively by verifying external tools.

The accepting approach has, for many years, been pursued in the PVS
ITP system, for instance by integrating the Yices SMT solver
\cite{Rushby06}. However, this is often insufficient for constructive
logic as proofs have computational content and may require execution.

The autarkic approach is the ideal, as an internally verified solver
is guaranteed to produce correct output. The \texttt{omega},
\texttt{tauto} and \texttt{ring} tactics in Coq, and Isabelle's
\textsf{blast} and \textsf{metis} tactics for instance, are
autarkic. The disadvantage of this approach however is clear:
there is a need to efficiently re-implement provers in the proof
system.

The approach taken in this paper approach is skeptical. We believe
this yields an adequate balance between efficiency and trust. Our
approach is heavily inspired by Isabelle's Sledgehammer tool, which
however is predominantly based on macro-step proof
reconstruction. Additionally, ATP integration in Mizar---so far
without proof reconstruction---is currently under development
\cite{RudnickiUrban11}. The skeptical approach has also been used in
the context of dependent types, in a Waldmeister integration into Agda
\cite{FosterStruth11}. The relative inefficiency of this approach due
to Agda proof normalisation is another main inspiration for
\sysname. Work on \emph{proof irrelevance} in the most recent version
of Agda, may however lead to a solution to this problem within Agda.
More recently, using the skeptical approach, an SMT solver has been
integrated into Coq~\cite{armand11coqsmt}.


\section{Conclusion and Future Work}

We have integrated the equational theorem prover Waldmeister into the
prototypical dependently typed programming language \sysname\ which is
based on the extended calculus of constructions with universes. In
contrast to previous approaches, where theorem provers were added a
posteriori to existing ITP systems complement existing internal
tactics and proof strategies, we take the ATP system as a core proof
engine for the programming language and build the language around
it. As a user front end we have implemented a proof scripting language
in the Proof General environment. This provides an interface between
\sysname\ and Waldmeister. Since Waldmeister provides highly detailed
proof output we can perform micro-step proof reconstruction,
translating the proof output into a \sysname\ proof term and type
checking that term.

Proof terms in \sysname\ are not normalised. On the one hand, this
makes proof reconstruction much more efficient. On the other hand this
yields a proof certificate rather than a proper normalised proof. The
strong normalisation property of the underlying type system, however,
guarantees that all proofs that have been successfully checked can
also be normalised. In the case of an equational proof this amounts to
a $\refl$ term.

In sum, our findings suggest that integrating ATP systems into DTP
languages can be very beneficial for program development in this
setting, and that the approach taken with \sysname\ may serve as a
template for future approaches to integrate more expressive ATP
systems in more sophisticated DTP languages.

There are various interesting directions for future work.

First, already the minimalist formalism of \sysname\ without recursion
or data types requires proofs in full multi-sorted first-order
constructive logic with equations. However, state-of-the-art ATP
systems are essentially all based on classical first-order logic and
often do not support sorts. Our current Waldmeister integration deals
only with multi-sorted equational logic, a fragment where classical
and constructive reasoning coincide. While using classical ATP systems
for more expressive fragments of first-order logic, such as Harrop
formulae, is still possible, specific ATP systems for constructive or
intuitionistic logic should be designed for applications in DTP.

Second, many state-of-the-art ATP systems adhere to a common input
standard (TPTP), but many of them do not provide any detailed proof
output or use a proprietary format. Detailed proof output is often
perceived as detrimental to proof search efficiency. In the context of
DTP, however, its absence is detrimental to proof reconstruction. As
Sledgehammer shows, macro-step proof reconstruction, that is,
replaying proof search with an internally verified theorem prover, has
the disadvantage that many proofs provided by the external ATPs will
not be accepted by the ITP system. Our proof experiments show that
micro-step reconstruction of individual proofs steps is superior to
this approach, but it requires detailed ATP output. Proof
standardisation as in the TSTP project~\cite{SutcliffeZimmerSchulz04}
is a valuable step in this direction. While sheer proof power was the
main emphasis of ATP development in the past, applications in the
context of ITP systems requires this to be balanced with detailed
proof output and support for types.

Third, in its current version, \sysname\ still suffers from the fact
that Waldmeister proofs which introduce new constants or variables
cannot always be reconstructed. We could work around this by
reconstructing proofs as they are, with additional constants and
variables included, and proving that such reconstructed proofs are
equivalent to the desired proofs. The simple heuristics currently used
should further be refined to cover more proofs. Alternatively, when
heuristics fail, the presence of lemmas in Waldmeister proof outputs
allows local manual proof reconstruction. Often, reconstruction
failures are caused by a very small number of lemmas. These could be
replaced by metavariables so that the proof can be delegated to
users.  Thus, even when and ATP system cannot completely finish a
proof, it might still produce a number of simpler proof goals for the
user and at least simplify the global proof goal.

Fourth, \sysname\ needs to be extended with features found in more
sophisticated DTP and ITP tools. First we could extend \ccop\ with
data types, induction or $\Sigma$-types. Alternatively we could extend
the proof scripting language by adding more automation, or by
providing a more structured method of proof construction, similar to
Isar. Some features like induction might only require proof management
such as induction tactics, and would not affect the ATP integration,
while others, such as the addition of $\Sigma$-types seem to require
modifications to how ATP systems are integrated.


\begin{thebibliography}{10}

\bibitem{armand11coqsmt}
M.~Armand, G.~Faure, B.~Gr{\'e}goire, C.~Keller, L.~Th{\'e}ry, and B.~Werner.
\newblock A modular integration of {SAT}/{SMT} solvers to {C}oq through proof
  witnesses.
\newblock In J.~Jouannaud and Z.~Shao, editors, {\em Certified Programs and
  Proofs}, volume 7086 of {\em LNCS}, pages 135--150. Springer, 2011.

\bibitem{Aspinall00}
D.~Aspinall.
\newblock Proof general: A generic tool for proof development.
\newblock In S.~Graf and M.~I. Schwartzbach, editors, {\em TACAS 2000}, volume
  1785 of {\em LNCS}, pages 38--42. Springer, 2000.

\bibitem{Awodey:2009}
S.~Awodey and M.~A. Warren.
\newblock Homotopy theoretic models of identity types.
\newblock {\em Math. Proc. Camb. Phil. Soc.}, 146:45--55, 2009.

\bibitem{BachmairDershowitzPlaisted89}
L.~Bachmair, N.~Dershowitz, and D.~A. Plaisted.
\newblock Completion without failure.
\newblock In H.~Ait-Kaci and M.~Nivat, editors, {\em Resolution of Equations in
  Algebraic Structures}, pages 1--30. Academic Press, 1989.

\bibitem{Barendregt:1991}
H.~Barendregt.
\newblock {Introduction to generalized type systems}.
\newblock {\em Journal of functional programming}, 1(2):125--154, 1991.

\bibitem{Barendregt02}
H.~Barendregt and E.~Barendsen.
\newblock Autarkic computations in formal proofs.
\newblock {\em Journal of Automated Reasoning}, 28(3):321--336, 2002.

\bibitem{Bernardy:2010}
Jean-Philippe Bernardy, Patrik Jansson, and Ross Paterson.
\newblock Parametricity and dependent types.
\newblock {\em SIGPLAN Not.}, 45:345--356, September 2010.

\bibitem{bertot2004interactive}
Y.~Bertot and P.~Cast{\'e}ran.
\newblock {\em Interactive theorem proving and program development: Coq'Art:
  the calculus of inductive constructions}.
\newblock Springer, 2004.

\bibitem{BlanchetteBulwahnNipkow11}
J.~C. Blanchette, L.~Bulwahn, and T.~Nipkow.
\newblock Automatic proof and dis- proof in {I}sabelle/{HOL}.
\newblock In C.~Tinelli and V.~Sofronie-Stokkermans, editors, {\em FroCoS
  2011}, volume 6989 of {\em LNCS}, pages 12--27. Springer, 2011.

\bibitem{norell09agda}
A.~Bove, P.~Dybjer, and U.~Norell.
\newblock A brief overview of {Agda} - a functional language with dependent
  types.
\newblock In Berghofer S., T.~Nipkow, C.~Urban, and M.~Wenzel, editors, {\em
  TPHOLs 2009}, volume 5674 of {\em LNCS}, pages 73--78. Springer, 2009.

\bibitem{Chargueraud:2011}
A.~Chargu\'{e}raud.
\newblock The locally nameless representation.
\newblock {\em Journal of Automated Reasoning}, 2011.
\newblock DOI 10.1007/s10817-011-9225-2.

\bibitem{Dybjer:1994}
P.~Dybjer.
\newblock Inductive families.
\newblock {\em Formal Aspects of Computing}, 6:440--465, 1994.

\bibitem{FosterStruth11}
S.~Foster and G.~Struth.
\newblock Integrating an automated theorem prover into {A}gda.
\newblock In M.~Bobaru, K.~Havelund, G.~Holzmann, and R.~Joshi, editors, {\em
  NASA Formal Methods}, volume 6617 of {\em LNCS}, pages 116--130. Springer,
  2011.

\bibitem{Nuprl}
PRL Group.
\newblock {\em Implementing Mathematics with the {N}uprl Proof Develop- ment
  System}.
\newblock Computer Science Department, Cornell University, 1995.
\newblock http://www.cs.cornell.edu/info/projects/nuprl/book/doc.html.

\bibitem{hillenbrand97waldmeister}
T.~Hillenbrand, A.~Buch, R.~Vogt, and B.~L\"{o}chner.
\newblock {Wald\-mei\-ster}: High performance equational deduction.
\newblock {\em Journal of Automated Reasoning}, 18(2):265--270, 1997.

\bibitem{Hurd05}
J.~Hurd.
\newblock System description: The {M}etis proof tactic.
\newblock In C.~Benzm{\"u}ller, J.~Harrison, and D.~Sch{\"u}rmann, editors,
  {\em ESHOL 2005}, pages 103--104. arXiv.org, 2005.

\bibitem{KnuthBendix70}
D.~Knuth and P.~Bendix.
\newblock Simple word problems in universal algebras.
\newblock In J.~Leech, editor, {\em Computational Problems in Abstract
  Algebra}, pages 263--297. Pergamon Press, 1970.

\bibitem{McBride04}
C.~McBride.
\newblock Epigram: Practical programming with dependent types.
\newblock In V.~Vene and T.~Uustalu, editors, {\em Advanced Functional
  Programming}, volume 3622 of {\em LNCS}, pages 130--170. Springer, 2004.

\bibitem{Miquel:2001}
A.~Miquel.
\newblock Le calcul des constructions implicite: syntaxe et s{\'e}mantique.
\newblock {\em These de doctorat, Universit{\'e} Paris}, 7, 2001.

\bibitem{Nordstrom:1990}
B.~Nordstrom, K.~Petersson, and J.~M. Smith.
\newblock {\em {Programming in Martin-L\"{o}f's Type Theory: An Introduction}}.
\newblock Oxford University Press, USA, 1990.

\bibitem{Tut:Norell:2008}
U.~Norell.
\newblock Dependently typed programming in agda.
\newblock In P.~W.~M. Koopman, R.~Plasmeijer, and D.~Swierstra, editors, {\em
  AFP'08}, LNCS, pages 230--266. Springer, 2009.

\bibitem{Pierce:2005}
B.C. Pierce.
\newblock {\em {Advanced topics in types and programming languages}}.
\newblock The MIT Press, 2005.

\bibitem{Pierce:2002}
B.C. Pierce~(Editor).
\newblock {\em {Types and programming languages}}.
\newblock The MIT Press, 2002.

\bibitem{RudnickiUrban11}
P.~Rudnicki and J.~Urban.
\newblock Escape to {ATP} in {M}izar.
\newblock PxTP 2011, 2011.

\bibitem{Rushby06}
J.~M. Rushby.
\newblock Tutorial: Automated formal methods with {PVS}, {SAL} and {Y}ices.
\newblock In D.~V. Hung and P.~Pandya, editors, {\em SEFM 2006}, page 262. IEEE
  Press, 2006.

\bibitem{SS98}
G.~Sutcliffe.
\newblock The {TPTP} problem library and associated infrastructure: The {FOF}
  and {CNF} parts, v3.5.0.
\newblock {\em Journal of Automated Reasoning}, 43(4):337--362, 2009.

\bibitem{SutcliffeZimmerSchulz04}
G.~Sutcliffe, J.~Zimmer, and S.~Schulz.
\newblock {TSTP} data-exchange formats for automated theorem proving tools.
\newblock In W.~Zhang and V.~Sorge, editors, {\em FroCoS 2004}, pages 201--215.
  IOS Press, 2004.

\end{thebibliography}

\end{document}